\documentstyle[epsf,emulateapj,pstricks]{article}

\def\exosat     {{\em EXOSAT}\/}
\def\asca       {{\em ASCA}\/}
\def\rosat      {{\em ROSAT}\/}
\def\am         {$^\prime$}
\def\deg        {$^{\circ}$}
\def\muller     {M\"{u}ller}
\def\kmsmpc     {~km$\;$s$^{-1}\,$Mpc$^{-1}$}
\def\ergs       {~erg$\;$s$^{-1}$}
\def\msun       {~$M_{\odot}$}
\def\msunyr     {$M_{\odot}\;$yr$^{-1}$}

\begin{document}

\lefthead{HEATING OF THE GAS IN TRIANGULUM AUSTRALIS}
\righthead{MARKEVITCH, SARAZIN \& IRWIN}

\title{HEATING OF THE INTRACLUSTER GAS IN THE TRIANGULUM AUSTRALIS CLUSTER}

\author{Maxim L. Markevitch\altaffilmark{1}, Craig L. Sarazin, Jimmy A.
Irwin}

\affil{Astronomy Department, University of Virginia, Charlottesville, VA
22903; mlm5y, cls7i, jai7e@virginia.edu}

\altaffiltext{1}{Also IKI, Profsoyuznaya 84/32, Moscow 117810, Russia}

\centerline{\small\it To appear in ApJ Letters, 1996 November 20}

\begin{abstract}

\asca\ and \rosat\ X-ray data are used to obtain two-dimensional maps of
the gas temperature, pressure and specific entropy in the Triangulum
Australis cluster of galaxies. We find that this hot ($T_e=10.3\pm0.8$ keV)
system probably has a temperature peak ($T_e>12$ keV) at the cluster core,
which approximately corresponds to the adiabatic relation. An underdense gas
sector, found in the \rosat\ cluster image eastward of the core, has a
higher temperature than average at that radius. At this higher temperature,
the gas pressure in this region is equal to that of the rest of the cluster
at the same radius, but the specific entropy of this gas is significantly
higher (although the temperature difference itself is only marginally
significant). We speculate that the existence of this region of underdense
high-entropy gas, as well as the adiabatic central temperature peak,
indicate recent or ongoing heating of the intergalactic medium in this
cluster. The most probable source of such heating is a subcluster merger,
for which the hydrodynamic simulations predict a qualitatively similar
temperature structure. We point out that entropy maps can provide a
physically meaningful way of diagnosing merging clusters and comparing the
predictions of merger simulations to the data.

\end{abstract}


\section{INTRODUCTION}

Spatially resolved measurements of the gas temperature in clusters of
galaxies can provide valuable information on the dynamical history of these
systems, pointing to those clusters with recent or ongoing merger activity
(see, for example, hydrodynamic merger simulations by Schindler \& \muller\
1993; Roettiger, Burns, \& Loken 1993; Evrard, Metzler, \& Navarro 1996).
Such measurements for hot clusters have become possible with the advent of
\asca\ with its combination of angular and spectral resolution and the wide
energy band of 0.5--10 keV (Tanaka, Inoue, \& Holt 1994). Two-dimensional
gas temperature maps of several clusters obtained with \asca\ using
independent techniques were presented, e.g., in Arnaud et al.\ (1994),
Markevitch et al.\ (1994), Markevitch (1996), Churazov et al.\ (1996), Honda
et al.\ (1996), Markevitch \& Vikhlinin (1996), and Henriksen \& Markevitch
(1996). The latter authors detected a prominent asymmetric temperature
pattern in the cluster A754, confirming with improved accuracy a similar
\rosat\ finding of Henry \& Briel (1995). The temperature nonuniformity in
A754, together with its complex X-ray image, indicated that it is undergoing
a major merger. A temperature structure of this kind was also found in
A3558, as well as in Coma. On the other hand, no significant asymmetric
features were detected in other well-resolved clusters, A2256, A2319 and
AWM7, except for the lower temperatures at the positions of the infalling
subunits in the former two systems. A significant radial temperature decline
at large radii was found with \asca\ in several clusters; for some of them,
most notably A2256, \rosat\ PSPC provided an independent confirmation
(refs.\ above).

Here we report on the detection of a possible merger signature in another
cluster, Triangulum Australis. It is a relatively nearby ($z=0.051$) bright,
hot system, which was overlooked in the optical band due to its low Galactic
latitude. It was first discovered as an X-ray source (McHardy et al.\ 1981).
From an \exosat\ observation, Edge \& Stewart (1991) obtained an average
temperature\footnote{Quoted errors are 90\% one-parameter intervals
throughout the paper.} of $8.0^{+1.4}_{-1.3}$ keV and an absorption column
of $2.5^{+1.8}_{-1.0}\times 10^{21}$ cm$^{-2}$, somewhat higher than the
Galactic value of $1.25\times 10^{21}$ cm$^{-2}$ (Heiles \& Cleary 1979).
The cluster has since been observed with \rosat, which produced a
high-quality image and an accurate measurement of the absorption column, and
with \asca, which provided a gas temperature map. These results are
presented below. At the cluster's redshift, $1\,h^{-1}$~Mpc $=25'\!.4$,
using $H_0\equiv 100\,h$ \kmsmpc.

\section{\rosat\ RESULTS}

The cluster was observed with \rosat\ PSPC for 7 ks in September 1992. We
have used the method and code of Snowden et al.\ (1994) to obtain the
flat-fielded images of the source in the energy bands R2--R7
(0.20--0.42--0.52--0.70--0.91--1.32--2.01 keV), free of non-cosmic
background. The cosmic component of the background was then calculated
individually in each energy interval using the outer parts of these images,
with the assumed accuracy of 5\% (Markevitch \& Vikhlinin 1996). Fluxes from
these images (excluding the poorly calibrated band R3) were fitted to
estimate the cluster temperature and the absorption column. We did not
include estimates of the PSPC calibration uncertainty to our analysis, since
our primary interest in the \rosat\ data is to obtain the image and the
absorption column value, well-determined by PSPC. Our best-fit $N_H$ value
is $1.42_{-0.25}^{+0.21}\times10^{21}$ cm$^{-2}$, and $T_e=7$ (4--17) keV,
the large temperature error being due to the high $N_H$. The absorption
column is consistent with the Galactic value and is marginally allowed by
\exosat. We also attempted to constrain the spectrum in the inner
$r=1'\!.5$ circle (half the core radius) of the cluster, to see if there is
any trace of a cooling flow to which PSPC is highly sensitive. We
approximately corrected for the energy dependence of the PSPC PSF (Hasinger
et al.\ 1992), which for this small radius makes a noticeable difference in
the spectral fit. Fixing $N_H$ at its value for the whole cluster and using
only 0.5--2 keV data, we obtain for this region $T_e=8.2$ (4.2--$\infty$)
keV. Freeing $N_H$ and including lower energies, the temperature is
$T_e$=5.0 (2.4--$\infty$) keV and $N_H=1.7\pm0.5\times10^{21}$ cm$^{-2}$.
Inclusion of a cooling flow component does not improve the fit
significantly; the best-fit inflow rate is 7 (0--33)~$h^{-2}$ \msunyr\ with
a similar best-fit absorption. The cluster image does not have a brightness
peak characteristic of a strong cooling flow, and our \asca\ spectra (below)
do not require a cooling flow component either. We conclude that there is no
indication of a significant cooling flow in the center.

The cluster image in the 0.5--2 keV band is presented by contours in Fig.~1.
Its X-ray brightness peak has coordinates $\alpha=16^{\rm h}38^{\rm
m}20^{\rm s}$, $\delta=-64^\circ21'22''$ (J2000). Assuming $T_e$=10 keV and
an iron abundance of 0.26 (average \asca\ values, see below) and the
absorption column derived above, the 0.5--2 keV cluster luminosity within
$r=1\,h^{-1}$ Mpc is $2.1\pm0.1\times 10^{44}\,h^{-2}$\ergs. Fitting the
brightness profile in the range of radii 0--25\am\ by a $\beta$-model of the
form $I_x(r)\propto (1+r^2/a_x^2)^{-3\beta+1/2}$ (Jones \& Forman 1984) and
ignoring the asymmetry of the image, we obtained $a_x=3'\!.5\pm0'\!.2$
($0.143\,h^{-1}$ Mpc) and $\beta=0.63\pm0.02$. We will use the derived $N_H$
value and the \rosat\ image in our \asca\ analysis below. Note the apparent
underdense sector east of the cluster center, which will be discussed below.
At the off-center distances of 3--10\am, the surface brightness in this
sector is about 2 times lower than that at the same radius on the opposite
side of the cluster.

\section{\asca\ ANALYSIS}

\asca\ observed Triangulum Australis in March 1995 in two pointings with
11.3 ks and 6.7 ks GIS useful time, which were offset by 7\am\ from one
another. The SIS has low sensitivity at high energies relevant for this hot
cluster, and was operated in 1-CCD mode which covers only
11\am$\times$11\am\ area, so we limited most of our analysis to the GIS
data. Using the energy band 1--11 keV, fixing the absorption column at the
\rosat\ best-fit value and fitting both pointings simultaneously, we
obtained an average cluster temperature of $10.3\pm0.8$ keV and an iron
abundance of $0.26\pm0.10$ (relative to Allen 1973). The \exosat\
temperature value ($8.0^{+1.4}_{-1.3}$ keV) is somewhat lower; however,
assuming their higher best-fit value of $N_H$, our temperature becomes 8.9
keV, consistent with \exosat.

For the reconstruction of the two-dimensional map of the cluster gas
temperature, the method described in detail in Markevitch (1996, and
references therein) has been used. It consists of simultaneous fitting of
the temperature in all image regions, using a 0.5--2 keV \rosat\ PSPC image
as a brightness template and modeling the \asca\ mirror scattering. We
restricted our spatially resolved analysis to the 2.5--11 keV energy band
due to the poor PSF calibration at lower energies. (A temperature value for
the whole cluster obtained from this energy band is $10.8_{-1.1}^{+1.3}$
keV.) The estimated systematic uncertainties of the background normalization
(20\%, $1\sigma$), the PSF model's wings (15\%) and core (5\%) and the
\asca\ effective area (5\%) were taken into account, as well as the
uncertainty due to the relative displacement of the \asca\ and \rosat\
images and the statistical error of the \rosat\ image. The nearest bright
X-ray source to Triangulum Australis is 3\deg\ away, so the possibility of
stray light contamination is excluded (Ishisaki 1996).

The image was divided into 3 concentric annuli with $r=0-3-12-23'$ centered
on the brightness peak. The second and third annuli were divided into 5 and
4 sectors, respectively, with one of the sectors coinciding with the
low-brightness region. The spectra from each region, each pointing and each
detector were binned in 7-10 energy intervals for adequate Gaussian
statistics. A uniform temperature over each image region was assumed. We
obtained $\chi^2_{\rm min}=149$ for 224 d.o.f.\ for the simultaneous fit.
The low reduced $\chi^2$ indicates that our conservative compound errors
are, if anything, slightly overestimated. The confidence intervals were then
estimated by Monte-Carlo simulations, including systematic errors. Figures 1
and 2{\em a} show the resulting temperature map. Two sectors to the north
and west of the outer ring which were poorly covered by \asca\ pointings are
not shown, but they were accounted for in the fit for completeness. There
are two features in the map which will be discussed below --- a significant
temperature peak in the cluster core, and a higher temperature in the
cluster low-brightness sector 2 compared to other regions at this radius.
(The contributions to the total $\chi^2_{\rm min}$ due to these regions are
24 and 14 for 28 data bins, respectively.) The latter temperature difference
by itself is statistically insignificant; however, due to the coincidence
with the depression of the gas density, the region's specific entropy
(Fig.~2{\em c}) is significantly different from the average at this radius.
The entropy values in sectors 2-6 could not be drawn from the same gaussian
distribution at the 95\% confidence, due to the single large deviation in
region 2.

We note that the \asca\ data alone do not require as large a central
temperature peak. As a consistency check, we performed a fit with the
emission measure of the central region (relative to other regions) being an
additional free parameter. Normally, the 0.5--2 keV relative brightness in
different regions is predetermined by the \rosat\ image in our technique.
(Note that the \rosat-\asca\ relative normalization is not used.) The
resulting best-fit relative brightness in the central annulus was 22\%
higher than that from \rosat, and the central temperature was lower, with
the temperatures in all other regions, as well as error bars, increasing
(dotted symbols in Fig.~1{\em a}). However, both quantities were consistent
within their 90--95\% confidence intervals with the temperature obtained
above and the \rosat\ brightness, respectively. Larger temperature errors in
such a fit are natural because the important piece of information provided
by the \rosat\ image is withdrawn. We have checked the possibility that
\rosat\ underestimates the relative emission measure in the $r<3'$ region by
20\%, for example, due to a greater absorption toward the center (a factor
of $>2$ increase is needed). The \rosat\ best-fit $N_H$ in the central
$r=3'$ region is $1.55_{-0.39}^{+0.35}\times10^{21}$ cm$^{-2}$, far short of
what is needed to produce such effect. The PSPC finite angular resolution
depletes the flux in the image peak only by $\sim 2$\%, which is absorbed by
the assumed \asca\ PSF uncertainty. A cooling flow, which would make the
\rosat\ image inadequate for our purpose, is not required by \rosat\
(above), nor does it make any improvement to either the GIS or the SIS fit.
(The short SIS exposure covering the central region does not provide useful
constraints for the main hot component.)

An optical image of Triangulum Australis shows that there is a cD galaxy
$0'\!.3$ from the X-ray brightness peak. If it has an active nucleus, it may
add a nonthermal component to the central flux. However, a 4.85 GHz radio
image (Condon et al.\ 1993) shows no source at the galaxy's position with an
upper limit of about 30 mJy (two orders of magnitude below the luminosity of
the cD galaxy NGC 1275 in the Perseus cluster). Therefore, a significant
nonthermal flux from this galaxy is unlikely, unless it is a radio-quiet
AGN, unusual in a cluster center. The best-fit normalization of an
additional X-ray component with a typical power-law AGN spectrum is zero,
although anything more exotic (for example, a heavily self-absorbed AGN with
$L_x$(2--10 keV)$\sim 2\times 10^{43}\,h^{-2}$\ergs) cannot be excluded, due
to the \asca's limited energy band.

We also tried to fit the relative brightness in sector 2 as another free
parameter, which practically did not change the temperature and the
normalization in this region compared to the value fixed by \rosat. Given
all the above, we will assume that the central discrepancy is probably a
statistical deviation, and our temperature measurement using the
\rosat\ image is correct, although it appears to be model-dependent. Our
major conclusions will hold even without the central temperature peak.

\section{DISCUSSION}

We will now discuss the implications of our temperature measurement.
Interestingly, from the \rosat\ image alone, which is rather circular except
for the low-brightness sector, one might expect that this sector has a
higher temperature, if the gas is in pressure equilibrium and the
gravitational potential of the cluster is reasonably smooth and symmetric.
Pressure variations not supported by gravity would disperse in roughly their
sound crossing time ($\sim 3\times 10^8\,h^{-1}$ yr for the low brightness
sector). Since this sector presumably contains lower density gas, it must
have a higher temperature to maintain a pressure similar to the adjoining
sectors. As our temperature map indicates, the gas in this region is indeed
likely to be hotter. It is useful to compare the gas pressure and specific
entropy in different regions, which is done in Figure 2({\em b,c}).  Here,
the entropy per particle is defined as $\Delta s\equiv
s-s_0=\frac{3}{2}k\,{\rm ln}\left[ (T/T_0)(\rho/\rho_0)^{-2/3}\right]$, the
subscript 0 denoting central values. The emission measure-weighted average
density in each sector was crudely estimated as the value from the symmetric
$\beta$-model distribution multiplied by the ratio of the square root of the
actual X-ray brightness to that in the model. Figure 2{\em b} indicates that
the gas pressure is nearly constant at a given radius. Thus, the gas could
be nearly in hydrostatic equilibrium with a symmetric gravitational
potential. Note, however, that the gas in region 2 is less dense than the
average in the middle annulus and would be expected to rise buoyantly on a
timescale of $\sim 10^9\,h^{-1}$ yr. Thus, the surface brightness
distribution in Figure 1 is not consistent with a stable equilibrium
distribution, and is likely to have resulted from a transient hydrodynamic
process. We note that the image can perhaps be explained as well by a very
irregular underlying mass configuration, but that would certainly mean that
the cluster is merging.

A given gravitational potential may support different gas configurations,
which are hydrostatic but at the same time differ by their histories and
their entropy distributions. The specific entropy $\Delta s$, shown in
Fig.~2{\em c} for our data, is a useful diagnostic of shock heating or of
any other nonadiabatic process. It may provide information on how the
particular, observed gas distribution has emerged. For comparison, Figure
2{\em c} also shows the entropy for an isothermal $\beta$-model (arbitrarily
choosing a temperature of 8 keV of the three cooler ``reference'' sectors of
the second annulus\footnote{We also chose the observed values of $a_x$ and
$\beta$ for clarity, although they would be different for a given binding
mass profile.}). This is the kind of distribution cluster gas is expected to
approach if left to itself, particularly if thermal conduction is effective.
Two gas regions significantly deviate from this isothermal equilibrium
model: the cluster core (if we accept that our central temperature is
correct) and sector 2. Another notable detail is that the gas in the core
has nearly the same specific entropy as the average in the surrounding
annulus. One explanation of this adiabatic temperature peak is that the core
gas has recently mixed turbulently with the surrounding gas. The higher
temperature in the core would then result from the adiabatic compression
required to bring the pressure in the core into hydrostatic equilibrium in a
given gravitational potential. The second annulus and the outer sectors have
similar temperatures and different entropies, which suggests that any such
mixing was confined to the central region. This is not unreasonable since
the core has only about 10\% of the gas within the outer boundary of our
second annulus.

However, the entropy in region 2 (the low brightness sector) is higher than
the average for that annulus or that of the cluster core. This could not be
due to mixing; it suggests that the gas in this region underwent a recent
heating event. For example, this might have been due to the passage of a
shock from a subcluster merger (e.g., Schindler \& M\"uller 1993; Pearce,
Thomas, \& Couchman 1994), or this gas may have been expelled from the core
where the heating occurred. Presumably, this gas then expanded adiabatically
into pressure equilibrium with the surrounding gas. Note that because the
pressure has already equalized but the entropy in this region is still high,
the event should have happened about $10^9\,h^{-1}$ yr ago. The same event
may have heated the central gas sufficiently to produce a radially
decreasing entropy gradient, after which convective mixing would act to
eliminate this gradient, producing a distribution with similar specific
entropy in the core and the surrounding gas. One can estimate the
amount of energy which should have been injected into the gas residing in
the core and region 2, assuming that the cluster was initially isothermal at
8 keV, as $Q \approx T \Delta S \sim (3-6)\times 10^{61}\,h^{-5/2}$~erg
(approximately equally distributed between the core and sector 2). To
estimate the heating which might occur in a merger, we consider the kinetic
energy acquired by the gas in two subclusters each with a total mass of
$6\times 10^{14}$\msun\ infalling from a distance of 5 to 1 Mpc; this energy
is about 10 times greater than required.  Therefore, a merger of subclusters
is a reasonable source of such heating. Hydrodynamic simulations predict
that a head-on merger of subclusters should produce a hot core and an
asymmetric temperature structure not unlike the observed one.

Other sources of heating might include supernova-driven galactic winds
(e.g., Evrard, Metzler, \& Navarro 1996). However, the present day supernova
rate expected for the stellar population in the core of this cluster is much
too small to produce enough heating on the timescale limited by the possible
lifetime of the observed temperature structure.

\section{SUMMARY}

Using \asca\ and \rosat\ data, we have reconstructed two-dimensional
distributions of the gas temperature, pressure and entropy in the Triangulum
Australis cluster. We detect a significant temperature peak in the cluster
core (although its existence is somewhat model-dependent), and a likely
temperature increase in the sector coincident with the underdense region of
the gas found in the cluster X-ray image. The underdense region is in
pressure equilibrium with the gas at the same radius but has a higher
specific entropy. We interpret the obtained entropy distribution as evidence
of the ongoing or recent heating of the intracluster medium, perhaps by
shocks from a subcluster merger.  A temperature map with higher angular
resolution to look into the structure of the cluster core, as well as
covering the outermost cluster regions, would be useful for understanding
the processes responsible for heating of the gas in this object, and perhaps
in clusters in general. The entropy distribution is a useful diagnostic of
shock heating or of any other nonadiabatic process. Therefore, it would be
very helpful if simulations of cluster hydrodynamics were analyzed to
produce the gas pressure and entropy maps, to enable a quantitative and
physically meaningful comparison to the data.

\acknowledgments

We thank the referee, A. C. Edge, for useful criticism. We gratefully
acknowledge the support from NASA grants NAG5-2526 and NAG5-1891.

\end{multicols}

\noindent
\begin{figure*}[h]
\pspicture(0,1)(8.9,20)

\rput[tl]{0}(-0.05,20.2){\epsfxsize=9.7cm
\epsffile{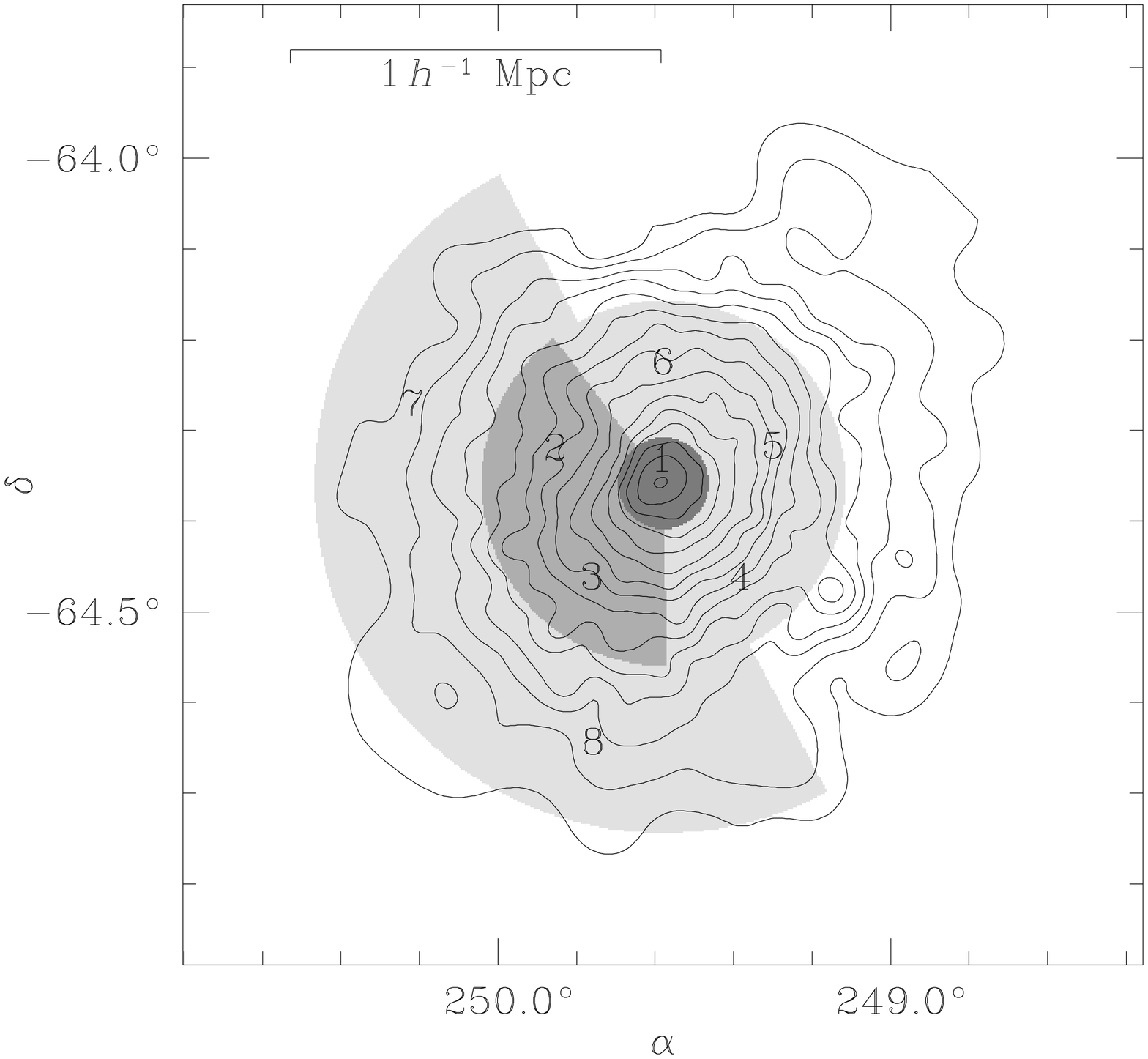}
}

\rput[tl]{0}(-0.1,11.3){
\begin{minipage}{8.85cm}
\small\parindent=3.5mm
{\sc Fig.}~1.---Temperature map of Triangulum Australis. Contours show the
\rosat\ PSPC surface brightness in the 0.5--2 keV band (smoothed with a
Gaussian of variable width for clarity). The contours are logarithmically
spaced with a factor of $\sqrt{2}$. The grayscale shows \asca\ temperatures
(darker is hotter). Regions (whole inner circle and two outer annuli divided
into 5 and 4 sectors, respectively) are numbered and their temperatures are
shown in Fig.~2{\em a}. The two sectors of the outer annulus which were
poorly covered by \asca\ are not shown.
\end{minipage}
}
\endpspicture
\pspicture(-0.75,1)(9,20)

\rput[tl]{0}(-0.1,20.1){\epsfxsize=9.0cm
\epsffile{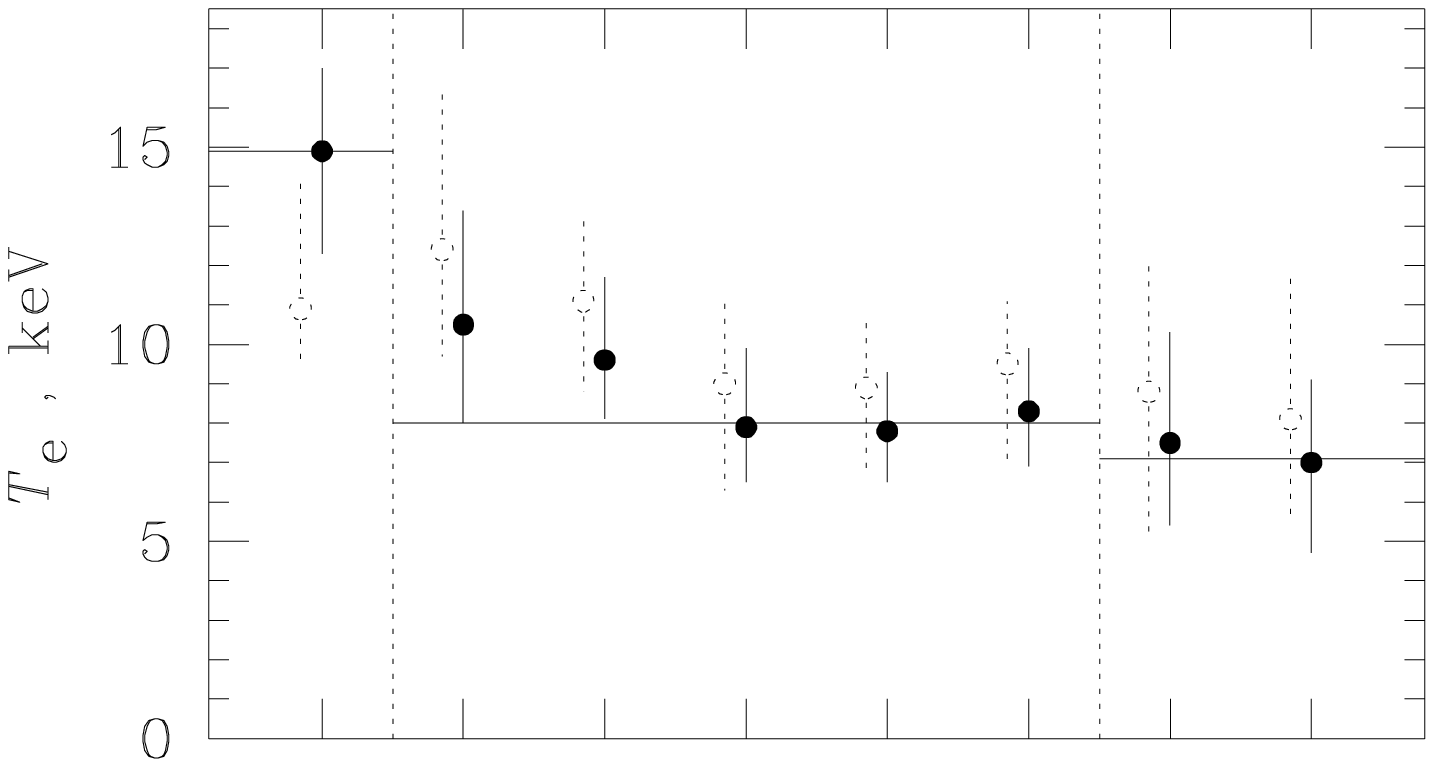}}

\rput[tl]{0}(-0.1,15.1){\epsfxsize=9.0cm
\epsffile{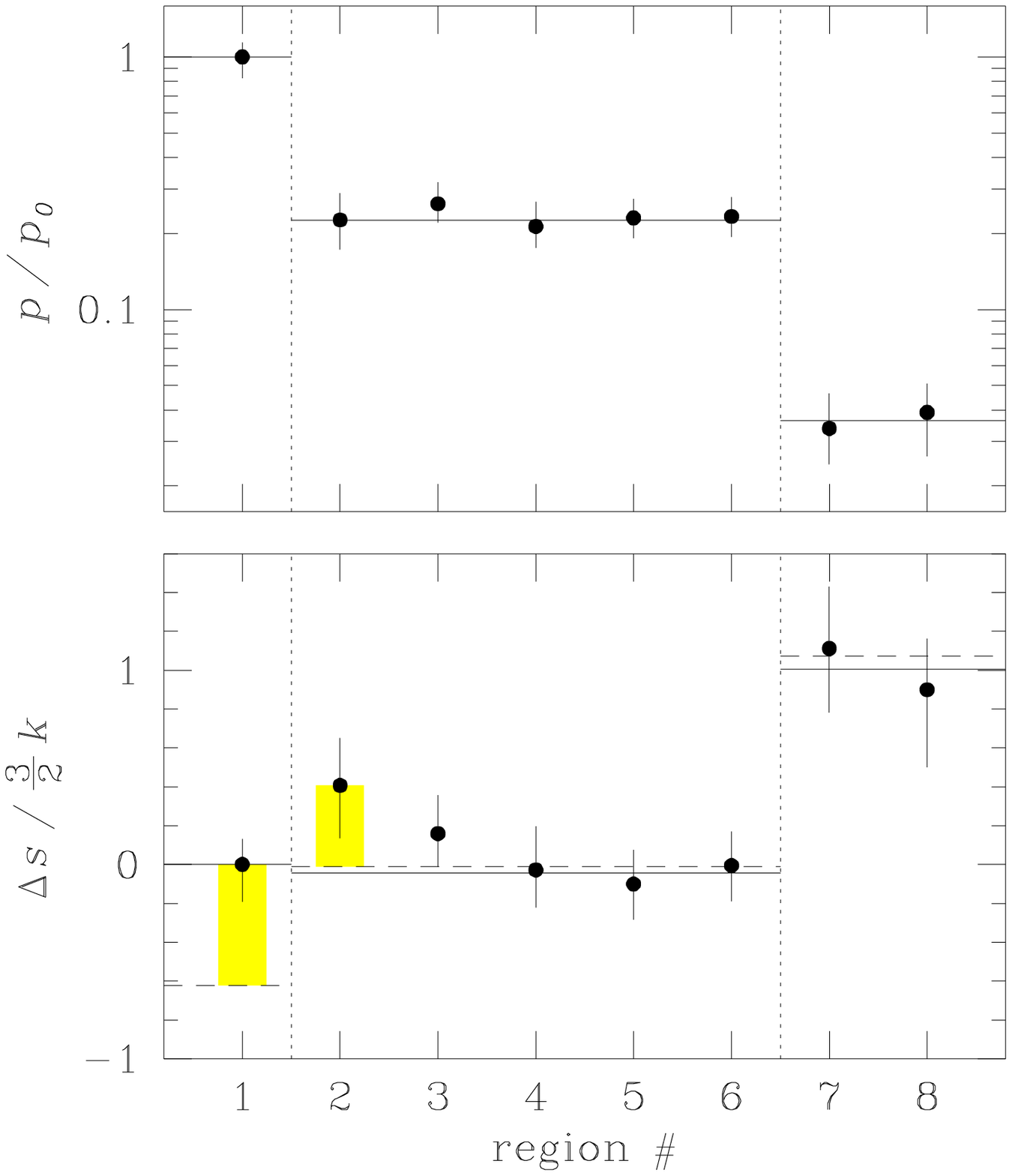}}

\rput[bl]{0}(8.4,19.5){\large\it a}
\rput[bl]{0}(8.4,14.52){\large\it b}
\rput[bl]{0}(8.42,9.7){\large\it c}

\rput[tl]{0}(-0.1,4.3){
\begin{minipage}{8.85cm}
\small\parindent=3.5mm
{\sc Fig.}~2.---Temperature ({\em a}), gas pressure ({\em b}) and specific
entropy relative to the central value ({\em c}) in the cluster regions
numbered in Fig.~1. Errors are 90\%. The dotted symbols in panel ({\em a})
correspond to a fit with the relative central brightness being a free
parameter (see the text). The solid horizontal lines correspond to the
average values over each annulus, excluding sectors 2 and 3 for the second
annulus. The dashed line in panel ({\em c}) corresponds to a symmetric
isothermal $\beta$-model with $T_e=8$ keV (the average temperature in our
``reference'' sectors 4, 5 and 6). The values of $\Delta s$ used for an
estimate of the amount of energy injected into the gas (see text) are shown
by shaded bars.
\end{minipage}
}
\endpspicture
\end{figure*}

\end{document}